\documentclass[pra,superscriptaddress,reprint,showpacs]{revtex4-1}
\usepackage{graphicx}
\usepackage{bm}
\usepackage{color}
\usepackage[USenglish]{babel}
\usepackage{amsmath}
\usepackage{multirow}
\newcommand{\mcsli}{\mathrm{CsLi}}
\newcommand{\mCsCs}{\mathrm{CsCs}}
\begin{document}
	
	
	\title{Role of the intraspecies scattering length in the Efimov scenario with large mass difference}
	
	\author{Stephan H\"afner}
	\author{Juris Ulmanis}
	\author{Eva D. Kuhnle}
	\affiliation{Physikalisches Institut, Universit\"at Heidelberg, Im Neuenheimer Feld 226, 69120 Heidelberg, Germany}
	
	\author{Yujun Wang}
	\protect\thanks{Present address: American Physical Society, 1 Research Road, Ridge, New York 11961, USA.}
	\affiliation{Department of Physics, Kansas State University, 116 Cardwell Hall, Manhattan, Kansas 66506, USA}

	\author{Chris H. Greene}
	\email{chgreene@purdue.edu}
	\affiliation{Department of Physics and Astronomy, Purdue University, West Lafayette, Indiana, 47907-2036, USA }
	
	\author{Matthias Weidem\"uller}
	\email{weidemueller@uni-heidelberg.de}
	\affiliation{Physikalisches Institut, Universit\"at Heidelberg, Im Neuenheimer Feld 226, 69120 Heidelberg, Germany}
	\affiliation{Hefei National Laboratory for Physical Sciences at the Microscale and Department of Modern Physics, and CAS Center for Excellence and Synergetic Innovation Center in Quantum Information and Quantum Physics, University of Science and Technology of China, Hefei, Anhui 230026, China} 
	
	\date{\today}
	
     \begin{abstract}
     		We experimentally and theoretically study the effect of the intraspecies scattering length onto the heteronuclear Efimov scenario, following up on our earlier observation of Efimov resonances in an ultracold Cs-Li mixture for negative [Pires \textit{et al.}, Phys. Rev. Lett. 112, 250404 (2014)] and positive Cs-Cs scattering length [Ulmanis \textit{et al.}, Phys. Rev. Lett. 117, 153201 (2016)]. Three theoretical models of increasing complexity are employed to quantify its influence on the scaling factor and the three-body parameter: a simple Born-Oppenheimer picture, a zero-range theory, and a spinless van der Waals model. These models are compared to Efimov resonances observed in an ultracold mixture of bosonic $^{133}$Cs and fermionic $^6$Li atoms close to two Cs-Li Feshbach resonances located at 843 G and 889 G, characterized by different sign and magnitude of the Cs-Cs interaction. By changing the sign and magnitude of the intraspecies scattering length different scaling behaviors of the three-body loss rate are identified, in qualitative agreement with theoretical predictions. The three-body loss rate is strongly influenced by the intraspecies scattering length.
     \end{abstract}
		
	
	\maketitle
	
\section{Introduction}

The Efimov scenario~\cite{Efimov1970,*Efimov1971,*Efimov1973,*Efimov1979} addressing universal properties of three particles interacting via resonant forces has become one of the cornerstones of modern few-body quantum physics~\cite{Braaten2006,Ferlaino2011,*Wang2013,*Frederico2012}. The hallmark of this bizarre effect is the manifestation of an infinite geometrical progression of three-body bound states, the Efimov states, that follow a discrete scaling law. Such series have been observed in experiments with homonuclear Bose~\cite{Huang2014}, three-component Fermi~\cite{Williams2009}, and heteronuclear Bose-Fermi~\cite{Pires2014,Tung2014,Ulmanis2016b} systems (see Ref.~\cite{Ulmanis2016a} for a recent review), to a good extent confirming one of the long-standing predictions: the universal law  $ a_-^{(n)} =\lambda a_-^{(n-1)} $,  where $  a_-^{(n)} $ is the position of the  $ n $th Efimov resonance that is described by the $ s $-wave scattering length~$ a $, and $ \lambda $ is the universal scaling factor that depends only on quantum statistics, number of resonant interactions, and mass ratios of the three atoms~\cite{Braaten2006,DIncao2006a}. The excited helium trimer $^4 $He$ _3$, as observed in a molecular beam experiment, has recently been found to accurately obey the predictions for a universal Efimov trimer~\cite{Kunitski2015}.

The universality in the Efimov scenario manifests itself at asymptotically large scattering lengths and thermal wavelengths. If the specific details of the short-range interactions are not resolved, their effect can be incorporated in a single quantity, the three-body parameter (3BP). In ultracold homonuclear systems that interact through pairwise $ -C_6/r^6 $ potentials the 3BP is connected to the molecular van der Waals (vdW) tails by an approximate, species-independent constant~\cite{Berninger2011,*Roy2013,*Wild2012,*Gross2009,*Gross2010,Huang2014a}. Apart from slight modifications due to short-range effects~\cite{Huang2014a,Huang2015a}, this general scaling constitutes an example of the so-called vdW universality~\cite{Wang2012,*Naidon2014} and can be seen as a precursor to a larger class of universal low-energy three-body observables that are governed by mutual power-law interactions at small particle separations~\cite{Wang2014a,Wang2015b,Naidon2014a}.

Although the Efimov effect in ultracold \emph{heteronuclear} gases is based on the same physical principles as in homonuclear systems, its signatures are much richer. The inclusion of an additional short-range length scale and interaction complicate the simple picture.  For a mass imbalanced system consisting of two heavy bosonic atoms that resonantly interact with one lighter atom, the 3BP is predicted to depend not only on the pairwise vdW tails, but also on the mass ratio and the intraspecies scattering length between the two heavy atoms~\cite{Wang2012d}. Furthermore, owing to denser trimer spectra~\cite{DIncao2006a} and a comparatively small $ |a_-^{(0)}|$, Efimov states are found in the transition regime between the long- and short-range dominated potential parts~\cite{Pires2014,Ulmanis2016}. Thus the spectra may contain contributions from both regimes. Additionally, in a real ultracold atomic system, the inter- and intraspecies interactions are generally not controlled independently and therefore deviations from the exact log-periodic scaling behavior are expected, reflecting in a period dependent scaling factor $\lambda^{(n)}$~\cite{DIncao2009a, Ulmanis2016}. The experimental exploration of such scenarios, however, so far has been limited. To date, mainly isolated Efimov resonances were observed, as for K-Rb~\cite{Hu2014,*Bloom2013,*Barontini2009,*Barontini2010} and $ ^7 $Li-$ ^{87} $Rb~\cite{Maier2015} mixtures. In the $ ^6$Li-$ ^{133}$Cs system successive Efimov resonances have been observed for attractive \cite{Pires2014,Tung2014, Ulmanis2016} and repulsive~\cite{Ulmanis2016b} CsCs interactions.
	
In this paper we juxtapose the heteronuclear Efimov scenario for two heavy bosons $ B $ and one distinguishable particle $  X $ for positive and negative intraspecies scattering lengths. We discuss three theoretical models at different levels of complexity that can be used to solve the three-body Schr\"odinger equation, and show that the boson-boson scattering length critically modifies the three-body energy spectrum. The qualitative influence of the boson-boson scattering length and the finite-range effects is already revealed with a minimalistic hybrid Born-Oppenheimer (BO) approximation, treating the $BX$ interaction as contact like and the $BB$ interaction as hardcore van der Waals potential. A more quantitative approach is given by the universal zero-range theory in the  hyperspherical adiabatic approximation~\cite{Ulmanis2016b}. This description also includes the universal $ BB $ dimer state for positive intraspecies interactions, which leads to a splitting of the traditional Efimov scenario into two Efimov branches. Efimov states with energy smaller than the dimer energy at the intraspecies unitarity do not connect to the three-body threshold and thus can lead to the absence of Efimov resonances in three-body recombination spectra. As the third model, we use the spinless van der Waals (vdW) theory, which models pairwise interactions with single-channel Lennard-Jones potentials~\cite{Wang2012d}. The inclusion of realistic finite-range potentials allows one to quantitatively compare the theoretical results with the experimentally determined three-body loss spectra of Cs+Cs+Li recombination, and extract Efimov resonance positions.

Finally, by comparing experimental three-body loss rates close to two different Cs-Li Feshbach resonances (FRs) we find two distinct scaling behaviors, in agreement with previous predictions~\cite{DIncao2009a}. Such distinctive scaling properties can be used, for example, to tune and significantly increase the three-body lifetime due to Cs-Cs-Li collisions, which is an important step towards studies of strongly interacting Bose-Fermi mixtures. 
	
This paper is structured as follows: In Sec.~\ref{sec:theory} we give an overview of the theoretical models to explain the measurements of three-body recombination spectra in a heavy-heavy-light system. The experimental procedure for the investigation of three-body recombination in the Cs-Cs-Li system is given in Sec.~\ref{sec:exp}. A comparison between experiment and theory for negative and positive intraspecies scattering length is given in Secs.~\ref{sec:L3neg} and \ref{sec:L3pos}, respectively. The scaling behavior of three-body recombination near overlapping Feshbach resonances is analyzed in Sec.~\ref{sec:powerlaws}.

\vspace*{0.1cm}
	
\section{Theoretical models}
\label{sec:theory}

The three-body wave function for two identical bosons $B$ with mass $m_{B}$ and one distinguishable particle $X$ with mass $m_X$ is determined by the three-body Schr\"odinger equation:

\begin{widetext}
\begin{equation}
\left[-\frac{\hbar^2}{m_B} \nabla^2_{\mathbf{r}}- \hbar^2\frac{2m_B+m_X}{4m_Bm_X} \nabla^2_{\mathbf{\rho}}+V_{BB}(r)+V_{BX}\left(\left|\mathbf{\rho}+\frac{\mathbf{r}}{2}\right|\right)+V_{BX}\left(\left|\mathbf{\rho}-\frac{\mathbf{r}}{2}\right|\right)\right]\Psi(\mathbf{\rho},\mathbf{r})=E\Psi(\mathbf{\rho},\mathbf{r}) ,
\label{eq:schroedinger}
\end{equation}
\end{widetext}
where $\hbar$ is the reduced Planck's constant, $\mathbf{r}$ the vector between the two bosons $B$, and $\mathbf{\rho}$ the vector from the center of mass of atoms $B$ to the $X$ atom. $V_{BB/BX}$ denote the intra- and interspecies interaction potentials, respectively. The corresponding scattering lengths are labeled $a_{BB}$ and $a_{BX}$.

Here, we employ three methods in order to solve the Schr\"odinger equation.

\subsection{Born-Oppenheimer approximation}

In order to get an intuitive understanding of the influence of the intraspecies scattering length and finite-range effects onto the heteronuclear Efimov effect we solve the three-body Schr\"odinger equation [Eq. \eqref{eq:schroedinger}] within the Born-Oppenheimer (BO) approximation for a large mass imbalance $m_X/m_B \ll 1$~\cite{Ulmanis2015b}. It is assumed that the motion of the light particle adapts almost immediately to the distance between the two heavy particles $\mathbf{r}$ and therefore the total wave function $\Psi(\mathbf{\rho},\mathbf{r})$ is considered to be separable,
\[
\Psi(\mathbf{\rho},\mathbf{r})=\psi(\mathbf{\rho};\mathbf{r}) \phi(\mathbf{r}),
\]
where $\phi(\mathbf{r})$ is the wave function of the two heavy particles and $\psi(\mathbf{\rho};\mathbf{r})$ the wave function of the light particle, which parametrically depends on $\mathbf{r}$. Within the BO approximation the Schr\"odinger equation can be separated into two coupled equations. By solving the equation for the light particle the BO potential is obtained and the three-body problem is reduced to an effective two-body problem. The energy spectrum can be found by solving the equation for the heavy particles,
\begin{equation}
\left[-\frac{\hbar^2}{m_B} \nabla^2_{\mathbf{r}} +V_{BB}(r)+E_{\mathbf{r}}\right] \phi(\mathbf{r})= E \phi(\mathbf{r}),
\label{eq:BO}
\end{equation}
where $E_{\mathbf{r}}$ is the BO potential, which is well known to be $E_{\mathbf{r}}=-\hbar^2 W(1)^2/2 m_X r^2$ for $r\ll |a_{BX}|$ and contact interaction between $B$ and $X$. Here $W(1)$ is the Lambert-W function which is connected to the Efimov scaling factor $s_0^2=m_B W(1)^2/2m_X -1/4$. We model the intraspecies interaction potential with a hard-core van der Waals potential,
\begin{equation*}
V_{BB}(r)=
\begin{cases}
\infty, & r<r_0,\\
-C_{6,BB}/r^6, &r>r_0,
\end{cases}\label{eq:vdW:C6modelPot}
\end{equation*}
where $C_{6,BB}$ is the dispersion coefficient which depends on the details of the electronic configurations. This allows one to introduce characteristic length scale $r_{\mathrm{vdW}}$ and energy scale $E_{\mathrm{vdW}}$ \cite{Chin2010}. The short-range cutoff $r_0$ is analytically connected to the scattering length $a_{BB}$~\cite{Gribakin1993,Flambaum1999}. We solve Eq. \eqref{eq:BO} numerically for the case of $a_{BX}\rightarrow \infty$ and $r> r_0$ and find the wave functions and energies of the trimer states.

\begin{figure}[t]
	\centering
	\includegraphics[width=1\linewidth]{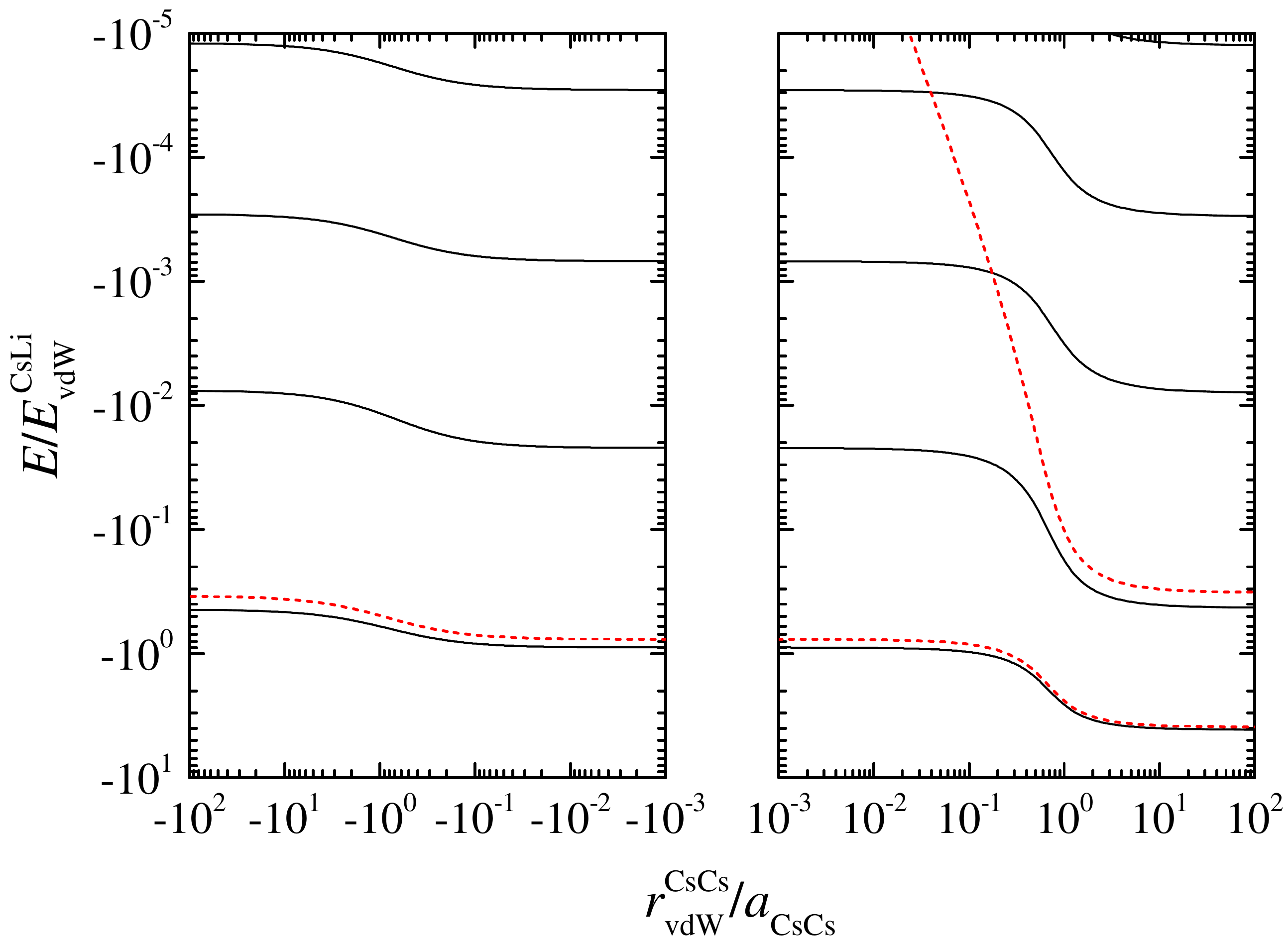}
	\caption{Born-Oppenheimer energy spectrum for the Cs-Cs-Li system vs the intraspecies scattering length $a_\mCsCs$. The spectrum is obtained for $a_\mcsli\rightarrow \infty$ (black solid lines). For comparison the energies of the two least bound Cs$_2$ states are displayed (red dashed lines).}
	\label{fig:BO}
\end{figure}

We apply the BO approximation to the case of the Cs-Cs-Li system with a mass ratio of $m_B/m_X=22.1$. The Cs-Cs scattering length $a_\mCsCs$ is changed by tuning the short-range cutoff $r_0$ of the intraspecies potential \cite{Gribakin1993}. The energy of four bound states is shown in Fig.~\ref{fig:BO} in dependence of  $a_\mCsCs$. The bound state energies (black lines in Fig.~\ref{fig:BO}) follow the discrete scaling law of the Efimov effect, but are strongly influenced by the intraspecies interaction and develop a steplike behavior around $|a_\mCsCs|\approx r_{\mathrm{vdw}}^{\mCsCs}$. For comparison we plot the two least bound states of the pure Cs$_2$ two-body hard-core vdW potential $V_{BB}(r)$. For $a\gg r_{\mathrm{vdW}}^{\mCsCs}$ the binding energy of the least bound state is given by the universal relation $E_b\propto -1/a_\mCsCs^2$. The binding energy of the second least bound state develops a gradual step at $a_\mCsCs\approx r_{\mathrm{vdW}}^{\mCsCs}$, due to its qualitative change from a vdW-dominated into a halo state (red dashed lines in Fig.~\ref{fig:BO}). Our BO model suggests a strong dependence of the Efimov state's energy and therefore also of the Efimov resonance positions $a_-^{(n)}$ at the three-body threshold on the intraspecies scattering length $a_\mCsCs$, which will be further studied within the universal zero-range theory and the spinless van der Waals theory.

\subsection{Zero-range theory}
\label{sec:zr}

For a qualitative description of the influence of the intraspecies scattering length between the two heavy bosons on the Efimov scenario, let us employ the universal zero-range theory in the hyperspherical adiabatic approximation, where pairwise contact interactions are assumed as used in \cite{Ulmanis2016b}. We introduce the relevant reduced mass factors:
\[
\mu _{BB}=\frac{m_B}{2},\text{ \ }\mu _{BX}=\frac{m_X m_B}{m_X+m_B},\text{ \ }\mu =%
\sqrt{\frac{m_X m_B^{2}}{2m_B+m_X}}.
\]
Relevant factors that relate to mass ratios are:%
\[
d_{BB}=\sqrt{\frac{m_X(2m_B)}{\mu (2m_B+m_X)}},\text{ \ }d_{BX}=\sqrt{\frac{m_B(m_X+m_B)}{%
		\mu (2m_B+m_X)}},
\]
and
\begin{eqnarray*}
\beta _{BB}&=&\arctan \left[ \sqrt{\frac{m_X(2m_B+m_X)}{m_B^{2}}}\right], \\
\beta _{BX}&=&\arctan \left[ \sqrt{\frac{(2m_B+m_X)}{m_X}}\right] .
\end{eqnarray*}
The following functions are also required as additional preliminaries,
before we can state the transcendental equation, that determines the adiabatic hyperspherical potentials, in a compact form:
\[
f(s,\alpha )\equiv \sqrt{\frac{2}{\cos \alpha }}\frac{\Gamma(\frac{3}{2})}{%
	\sin \alpha }P_{s-1/2,-1/2}(\sin \alpha ) ,
\]%
where $\Gamma(z)$ is the gamma function and $P_{n,m}$ is the associated Legendre function of the first kind and
\begin{eqnarray*}
W(s)&=& \frac{\pi \Gamma (\frac{2+s}{2})\Gamma (\frac{2-s}{2})}{2\Gamma (\frac{3}{2})^{2}\Gamma (\frac{1+s}{2})\Gamma (\frac{1-s}{2})}, \\
X(s)&=&\frac{\Gamma (\frac{2+s}{2})\Gamma (\frac{2-s}{2})}{2\Gamma (\frac{3}{2})^{2}}f(s,\beta _{BX}), \\
Y(s)&=&\frac{\Gamma (\frac{2+s}{2})\Gamma (\frac{2-s}{2})}{2\Gamma (\frac{3}{2})^{2}}f(s,\beta _{BB}).
\end{eqnarray*}
Then the roots $s(R),$ which determine the hyperradial potential curves
\[
U(R)=\frac{\hbar ^{2}}{2\mu }\frac{s(R)^{2}-\frac{1}{4}}{R^{2}}
\]%
are solutions of the following transcendental equation:%
\begin{widetext}
\[
\left( W(s)\frac{\mu a_{BB}}{\mu _{BB}d_{BB}^{3}}-R\right) \left( \frac{\mu a_{BX}}{%
	\mu _{BX}d_{BX}^{3}}(W(s)-Y(s))-R\right) -2X(s)^{2}\frac{\mu ^{2}a_{BX}a_{BB}}{\mu
	_{BX}d_{BX}^{3}\mu _{BB}d_{BB}^{3}}=0.
\]
\end{widetext}
Here $R$ is the hyperradius defined by
\[
\mu R^2=\mu_{BB} r^2+\frac{2m_B m_X}{2m_B+m_X}\rho^2 .
\]
Note that in general there could be real solutions $s(R)$ corresponding to
hyperspherical potential curve energies higher than $\frac{-\hbar ^{2}}{8\mu R^{2}}$, which is the critical coefficient to support an infinite number of three-body bound states, and imaginary solutions for $s(R)$ that correspond to potential curve energies below $\frac{-\hbar ^{2}}{8\mu R^{2}}.$

\begin{figure}[b]
	\centering
	\includegraphics[width=1\linewidth]{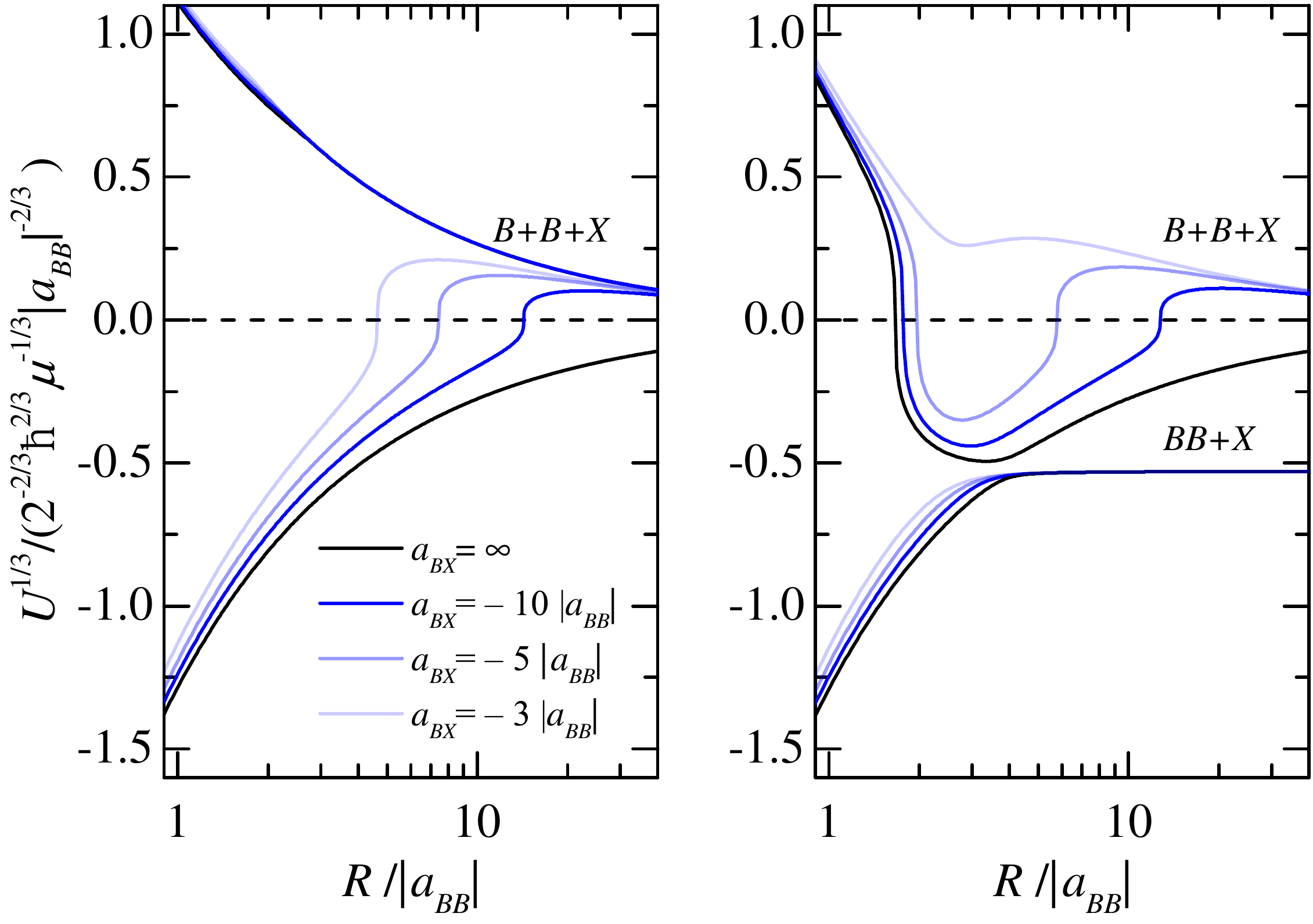}
	\caption[bound state energies]{Adiabatic hyperspherical potentials $ U $ for a $ BBX $ system with a mass ratio of $ m_B/m_X =22.1$ and $ a_{BB}<0$ (left panel) and $ a_{BB}>0 $ (right panel) in the universal zero-range model.  The curves correspond to interspecies scattering lengths $ a_{BX} $ with respective values $ \infty $, $ -10 $, $ -5 $, $ -3 $ in units of $ \left|a_{BB}\right| $. Figure taken from \cite{Ulmanis2016b}.
	}
	\label{fig:Cs2Li_Pot2}
\end{figure}

The heavy-heavy-light adiabatic hyperspherical potential curves are shown in Fig.~\ref{fig:Cs2Li_Pot2} for the case of positive and negative intraspecies scattering length $a_{BB}$. Here, a mass ratio of $m_B/m_X=22.1$ and different values of $a_{BX}<0$ are assumed. If both scattering lengths are negative, the original Efimov scenario with an $\propto -1/R^2$ hyperspherical potential is recovered for $a_{BX}\rightarrow \infty$, supporting an infinite number of bound states (Fig. \ref{fig:Cs2Li_Pot2}, left panel). For finite interspecies scattering lengths a potential barrier at $R\approx 2|a_{BX}|$ leads to quasi-bound states, generating three-body recombination resonances, when passing the dissociation threshold. In this case a three-body parameter is indispensable to regularize the energy spectrum.

For positive intraspecies scattering lengths ($a_{BB}>0$) the situation is considerably altered by the existence of a dimer $BB$ with binding energy $E_{BB}$ (Fig. \ref{fig:Cs2Li_Pot2}, right panel). The channel $BB+X$ splits the potential curves into two Efimov branches. The lower branch, which is asymptotically connected to the $BB+X$ channel, recovers the original Efimov scenario with a potential that is proportional to $-1/R^2$. In this case regularization with a three-body parameter is required to avoid a diverging binding energy of the ground state. Note that there is no potential barrier arising for finite values of $a_{BX}$. However, the upper Efimov branch with energies $E>E_{BB}$ shows a different behavior. It exhibits a potential barrier at $R\approx 2a_{BB}$, roughly independent of $a_{BX}$. Therefore, even for resonant interactions ($a_{BX}\rightarrow \infty$), where the $\propto -1/R^2$ potential is restored for large interpaticle separations, the energy spectrum is well defined and no regularization with a three-body parameter is necessary. For finite $a_{BX}$ an effective potential barrier forms at $R\approx 2|a_{BX}|$, leading to recombination resonances.

\subsection{Spinless van der Waals theory}

Additionally we solve the three-body problem within the spinless vdW theory \cite{Wang2012d}. The approach consists of numerically solving the three-body Schr\"odinger equation in the full hyperspherical formalism, where the two-body interaction potentials $V_{BB/BX}$ between equal bosons and the third particle separated by distance $r_{BB/BX}$ are modeled by single-channel Lennard-Jones potentials with vdW tails.
\[
V_{BB/BX}(r_{BB/BX})=-\frac{C_{6,BB/BX}}{r_{BB/BX}^6}\left[1-\left(\frac{r_{c,BB/BX}}{r_{BB/BX}}\right)^6\right] ,
\]
where $C_{6,BB/BX}$ are the dispersion coefficients. The scattering lengths $a_{BX}$ and $a_{BB}$ are reproduced by tuning the short-range cutoffs $r_{c,BB/BX}$. We assume that the hyperradial and hyperangular motions are approximately separable \cite{Macek1968} and treat the hyperradius as adiabatic parameter. The Schr\"odinger equation is reduced to a set of coupled 1D equations \cite{Wang2010c}. We include the nonadiabatic couplings between the hyperspherical potentials and assume a $J^\Pi=0^+$ symmetry \footnote{The absence of the Efimov effect in higher partial waves for our mass ratio of $m_B/m_X=22.1$ \cite{DIncao2006}, leads to a substantial suppression of $J>0$ contributions to the recombination rate in the ultracold regime with $k_B T\ll E_{\mathrm{vdW}}$. The suppression is estimated to be $(k r_\mathrm{vdW})^{2p_0} \ll 1$ for $a_\mCsCs<0$ and $(k a_\mCsCs)^{2p_0}\ll 1$ for the $a_\mCsCs\approx+190a_0$ case~\cite{Wang2011d}, where $p_0>1$ is a universal constant for the given mass ratio \cite{DIncao2006} and $k$ the thermal wavevector.} (the dominant contribution at ultracold temperatures), where $J$ is the total angular momentum and $\Pi$ the total parity.

\begin{figure}[t]
	\centering
	\includegraphics[width=1\linewidth]{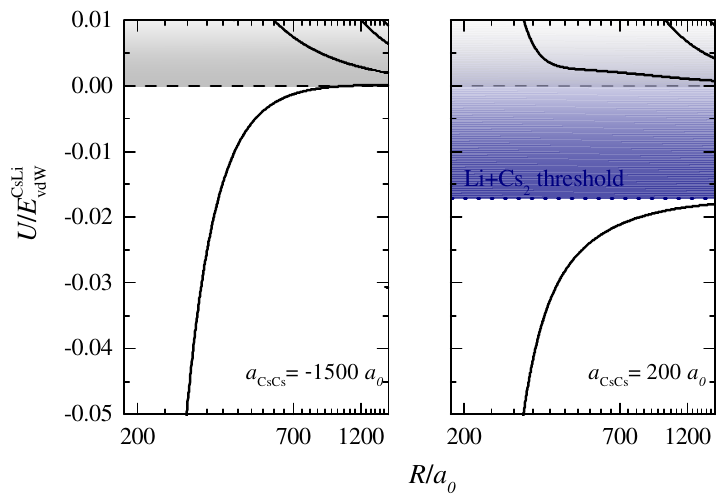}
	\caption[bound state energies]{Cs-Cs-Li adiabatic hyperspherical potentials $U$ within the spinless vdW theory for intraspecies scattering lenths $ a_{\mCsCs}=-1500a_0$ (left panel) and $ a_{\mCsCs}=+200a_0$ (right panel) in units of the Cs-Li vdW energy $E_{\mathrm{vdW}}^\mcsli$. In both cases the interspecies scattering length is $a_\mcsli=-400a_0$.}
	\label{fig:pot_vdw}
\end{figure}

This formalism is applied to the Cs-Cs-Li system and allows us to compare the theoretical predictions directly to our experimental observations. The interaction potentials are tuned such that the scattering lengths $a_\mcsli$, $a_\mCsCs$, and their functional dependence $a_\mCsCs(a_\mcsli)$ are reproduced for the experimentally employed field ranges (see Fig. \ref{fig:scatteringlengths}). The adiabatic hyperspherical potentials are plotted in Fig.~\ref{fig:pot_vdw} for intraspecies scattering lengths of $a_{\mCsCs}=-1500a_0$ and $a_{\mCsCs}=+200a_0$ resembling the two cases of the 843~G and 889~G Cs-Li FRs. The interspecies scattering length $a_{\mcsli}=-400a_0$ is tuned close to the value, where the ground-state Efimov resonance is expected~\cite{Pires2014}. For positive intraspecies interactions an avoided crossing originating from the coupling between the attractive three-body potential and a repulsive atom-dimer channel leads to an effectively repulsive Efimov potential. This prevents the scattering wavefunction from probing short-range parts of the potential and, consequently, recombination. Such an avoided crossing
is not present for the case of $a_{\mCsCs}=-1500a_0$, where the original Efimov effect is recovered.

The adiabatic potentials calculated by the spinless vdW theory closely resemble those by zero-range theory when $R$, $|a_{BB}|$, and $|a_{BX}|$ are all significantly greater than $r_{\mathrm{vdW}}^{\mCsCs}$ and $r_{\mathrm{vdW}}^{\mcsli}$. The universal condition $|a_{BB}|$, $|a_{BX}|$ $\gg$ $r_{\mathrm{vdW}}^{\mCsCs}$, and $r_{\mathrm{vdW}}^{\mcsli}$ is satisfied or approximately satisfied in both cases of the 843G and 889G Feshbach resonances; therefore, the zero-range theory is still helpful for understanding the scaling behavior of the three-body losses observed in our experiment, as discussed above. The quantitative determination of the observables, especially of the 3BP, requires the knowledge of the three-body dynamics when $R$ is smaller than or comparable with the van der Waals lengths, which needs the spinless van der Waals theory to resolve.

We calculate the energy-dependent three-body recombination rate from the S matrix \cite{Mehta2009,Wang2010c} and perform thermal averaging \cite{Suno2003}, assuming Boltzmann distributions with the experimentally determined temperatures. The width of the experimentally observed recombination features is reproduced by adding a heuristic hyperradial loss channel that assumes near unity loss at short distances, without modifying the resonance positions~\cite{Wang2015b}. The obtained three-body recombination rates are plotted in Figs. \ref{fig:loss_rate_energies}(a) and \ref{fig:loss_rate_energies}(c) together with the energy spectrum of the three energetically lowest Efimov states [see Figs. \ref{fig:loss_rate_energies}(b) and \ref{fig:loss_rate_energies}(d)].
For the case of the 843~G Cs-Li FR, where $a_\mCsCs \approx -1500a_0$, the classical Efimov scenario is recovered [Fig. \ref{fig:loss_rate_energies}(b)]. By lowering $a_\mcsli$ the Efimov states successively disappear by crossing the three-body scattering threshold. For positive intraspecies scattering length $a_\mCsCs\approx + 190a_0$, as it is the case for the 889~G Cs-Li FR, the Efimov sates split into two branches (Fig. \ref{fig:loss_rate_energies}(d)). While the lowest state predissociates into a Cs$_2$+Li state (blue dashed line) before reaching the three-body threshold and consequently does not lead to a recombination resonance at the three-body threshold, the higher-lying states recover the original Efimov scenario \cite{Ulmanis2016b}.

\section{Comparison with experiment}
\label{sec:L3}

\subsection{Cs-Li Feshbach resonances}

\begin{figure}[b]
	\centering
	\includegraphics[width=1\linewidth]{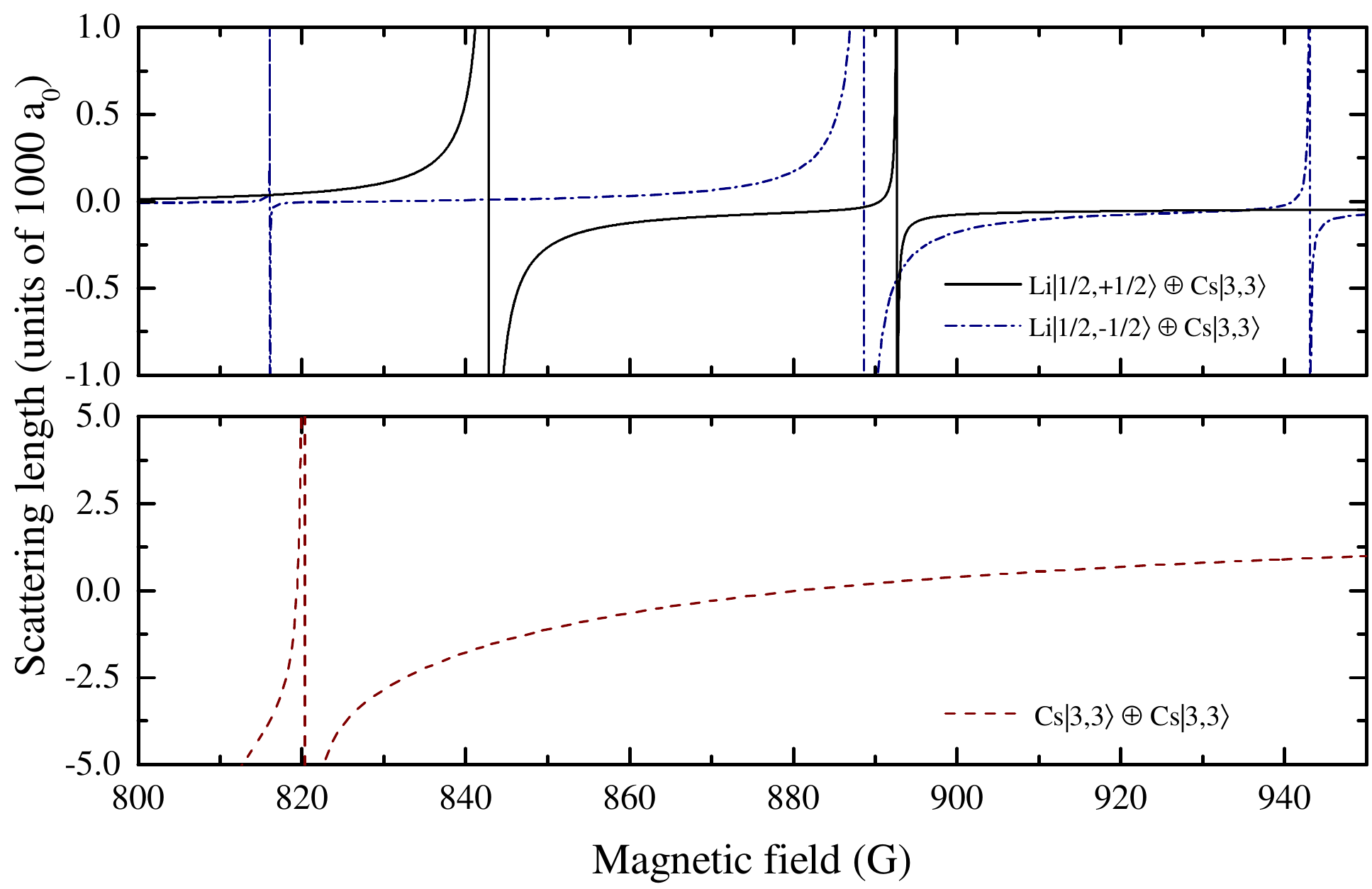}
	\caption[scattering lengths]{Interspecies scattering lengths $a_\mcsli$ for the channels Li$\left|1/2,+1/2\right\rangle\oplus\ $Cs$\left|3,3\right\rangle$ (upper panel, black line), Li$\left|1/2,-1/2\right\rangle \oplus\ $Cs$\left|3,3\right\rangle$ (upper panel, blue dash dotted line), and intraspecies scattering length $a_\mCsCs$ for Cs atoms in the ground state Cs$\left|3,3\right\rangle \oplus\ $Cs$\left|3,3\right\rangle$ (lower panel, red dashed line). Scattering lengths taken from \cite{Ulmanis2015,Berninger2013}.
	}
	\label{fig:scatteringlengths}
\end{figure}

The Cs-Cs-Li system offers the unique possibility to study the influence of the intraspecies scattering length onto the heteronuclear Efimov scenario. The experimentally adjustable interspecies and intraspecies scattering lengths $a_\mcsli$ and $a_\mCsCs$ are shown in Fig. \ref{fig:scatteringlengths}. Two intermediately broad ($s_{\mathrm{res}}\approx 0.7$~\cite{Tung2013}) Cs-Li Feshbach resonances (FRs), located at approx. 843~G and 889~G, are well suited for the study of the heteronuclear Efimov scenario \cite{Repp2013,Tung2013,Pires2014a,Ulmanis2015}. They are characterized by different sign and magnitude of the Cs-Cs scattering length $a_\mCsCs$: while close to the 843~G resonance the intraspecies scattering length is large and negative $a_\mCsCs\approx -1500a_0$, the 889~G resonance is characterized by a small and positive $a_\mCsCs\approx +190a_0$ \cite{Berninger2013}.
By investigating the narrow resonances located at approximately 816~G, 889~G, and 943~G characterized by $s_{\mathrm{res}}\leq 0.03$~\cite{Tung2013}, the Cs-Cs-Li system offers the possibility to investigate the influence of the multichannel nature of Feshbach resonances onto the universal Efimov scenario \cite{Wang2014a}, which has been studied recently~\cite{Johansen2017}.

\subsection{Experimental determination of $L_3$}
\label{sec:exp}

Our experimental procedure is described in detail in Refs. \cite{Pires2014, Ulmanis2016}. In brief we prepare an ultracold mixture of fermionic $^6$Li atoms in one of the two energetically lowest spin states $\left|f,m_f\right\rangle=\left|1/2,1/2\right\rangle$ or $\left|1/2,-1/2\right\rangle$ and bosonic $^{133}$Cs in the absolute ground state $\left|3,3\right\rangle$. Here $f$ and $m_f$ refer to the total angular momentum and its projection. By usage of a bichromatic trapping scheme \cite{Ulmanis2016} we prepare samples of $1\times 10^4$ ($7\times 10^3$) Cs (Li) atoms at temperatures as low as 120~nK. The spatial overlap of the two atomic clouds is approximately 45\% and assumed to be constant within the investigated magnetic field range. The measured trapping frequencies are $\omega_{\mathrm{Cs}}=2\pi\times (5.7, 115,85)$~Hz and $\omega_{\mathrm{Li}}=2\pi\times (25, 160,180)$~Hz.

The three-body recombination rates are measured in dependence of the external magnetic field analogous to our previous work \cite{Ulmanis2016,Ulmanis2016b}. We prepare the atomic mixture approximately 4~G away from the pole of the two broad interspecies Feshbach resonances at 843~G and 889~G. Within 150~ms we increase the dipole trap potential by 10\% in order to stop residual plain evaporation and to let the magnetic field stabilize. This leads to a temperature increase on the order of 10\%. The final magnetic field value is set by a fast ramp. After a variable hold time, both atomic species are imaged by high-field absorption imaging from which atom numbers and cloud sizes are deduced.

The three-body loss coefficient $L_3$ for the loss of one Li and two Cs atoms is retrieved by numerically fitting the coupled rate equations,
\begin{eqnarray}
\dot{N}_{\mathrm{Li}} &=& -\mathcal{L}_{1}^{\mathrm{Li}}{N}_{\mathrm{Li}} -\mathcal{L}_3 {N}_{\mathrm{Li}} {N}_{\mathrm{Cs}}^{2} ,
\label{lossrates1}
\\
\dot{N}_{\mathrm{Cs}} &=& -\mathcal{L}_{1}^{\mathrm{Cs}}{N}_{\mathrm{Cs}} - 2\mathcal{L}_3 {N}_{\mathrm{Li}} {N}_{\mathrm{Cs}}^{2} - \mathcal{L}_3^{\mathrm{Cs}}{N}_{\mathrm{Cs}}^{3} ,
\label{lossrates2}
\end{eqnarray}
where $\mathcal{L}_{1}^{\mathrm{Li}}$ and $\mathcal{L}_{1}^{\mathrm{Cs}}$ are the one-body loss rates for each species in the trap and $\mathcal{L}_3^{\mathrm{Cs}}$ the Cs three-body loss rate, which are determined in independent single-species measurements under the same experimental conditions. Hence $\mathcal{L}_3$ and the initial atom numbers ${N}_{\mathrm{0,Li}}$ and ${N}_{\mathrm{0,Cs}}$ are the only fitting parameters. The three-body loss rate coefficient $L_3$ is obtained from $\mathcal{L}_3$ via modeled atomic density distributions. The error bars are obtained by bootstrapping and resemble one standard deviation of the resampled distribution. The systematic error for the determination of the absolute value of $L_3$ is estimated to be a factor of 3 (120~nK data \cite{Ulmanis2016}) and 0.8 (450~nK data \cite{Pires2014}), respectively, and are mainly caused by uncertainties in the determination of atom cloud temperatures, densities, overlap, and trapping frequencies. Since no significant increase in temperature during the hold time is observed \cite{Ulmanis2016}, we neglect recombinational heating \cite{Weber2003} in our analysis.
The magnetic field stability is around 16~mG (one standard deviation) resulting from long-term drifts, residual field curvature along the long axis of the cigar-shaped trap, and calibration uncertainties. Due to intraspecies losses, the extraction of $L_3$ is limited to scattering lengths of $|a_\mcsli| \gtrsim 1000a_0$ for our data at 120~nK \cite{Ulmanis2016} and $|a_\mcsli| \gtrsim 400a_0$ for the data at the 889~G FR and temperature of 320~nK. The given loss rates in this region represent an upper bound of the actual $L_3$.

\subsection{Negative intraspecies scattering length}
\label{sec:L3neg}

	\begin{figure*}[t]
		\centering
		\includegraphics[width=1\textwidth]{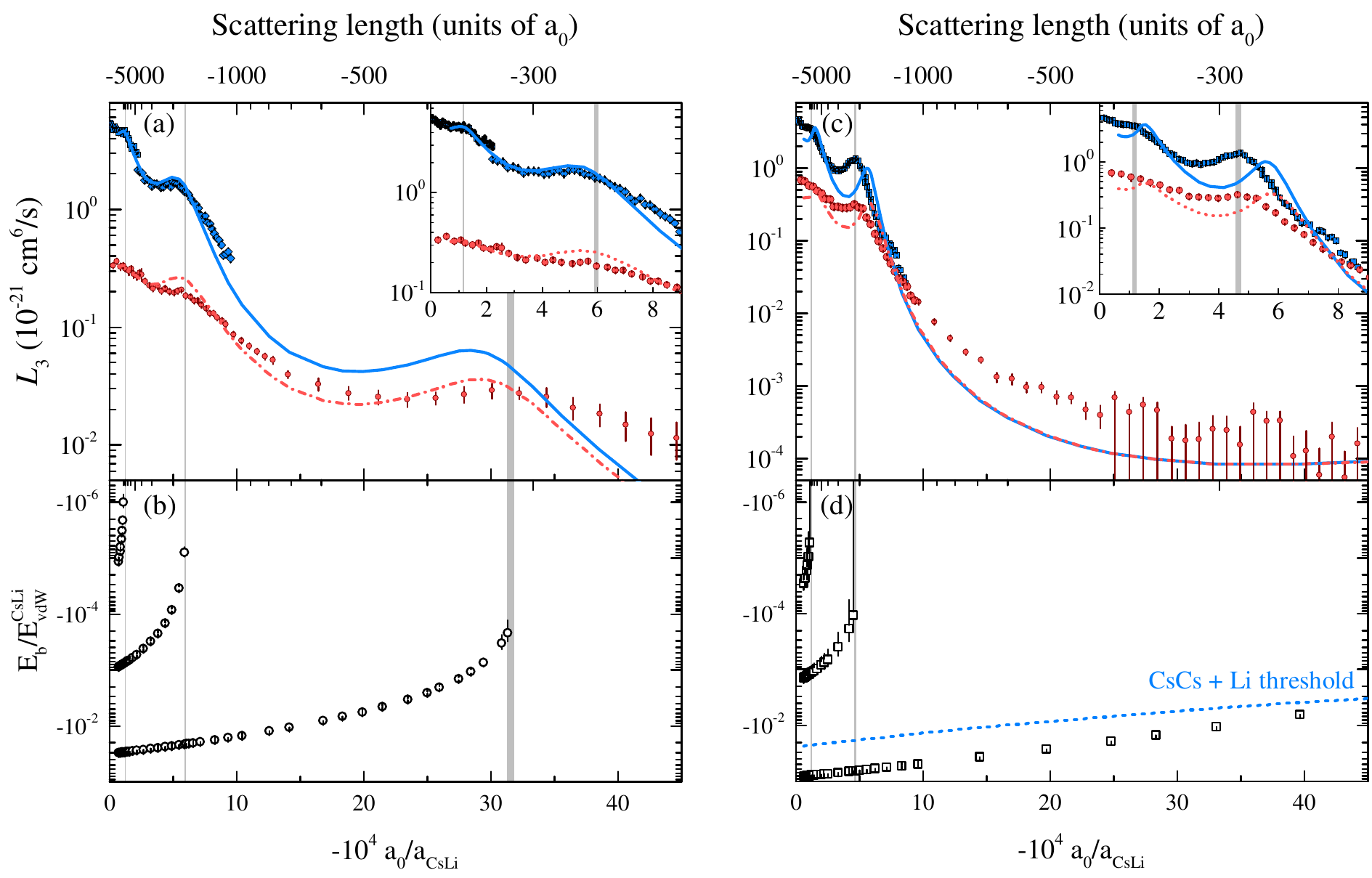}
		\caption[loss_rate_energies]{(a),(c) Cs-Cs-Li three-body recombination rate spectra at temperatures of (a) 120~nK (blue diamonds and squares, blue solid line) and 450~nK (red circles, red dash-dotted line) close to the 843~G Cs-Li Feshbach resonance and Cs-Cs scattering length $ a_\mCsCs\approx -1500a_0 $ and (c) 120~nK (blue squares, blue solid line) and 320~nK (red circles, red dash-dotted line) close to the 889~G Cs-Li Feshbach resonance and Cs-Cs scattering length $ a_\mCsCs\approx +190a_0 $. The experimental data are taken from \cite{Pires2014,Ulmanis2016,Ulmanis2016b}. The insets show a zoom into the region of the two excited Efimov resonances. The error bars represent statistical errors from bootstrapping, magnetic field uncertainties, and the technical limit due to Cs-Cs-Cs three-body losses. The blue solid lines and red dash-dotted lines show the calculated three-body loss rates from the spinless vdW theory for the lower and higher temperature, respectively. Experimental data have been scaled by a numerical constant, well within the absolute error, to fit the theory (see Ref.~\cite{Ulmanis2016}).
		(b),(d) Calculated CsCsLi energy spectra for the three deepest bound Efimov states for experimental scattering lengths that correspond to the (b) 843~G and (d) 889~G Cs-Li Feshbach resonances, respectively. The atom-dimer scattering threshold CsCs$ + $Li for the latter case is shown as a dashed blue line. The error bars represent the width of the corresponding Efimov state. Missing error bars indicate a width that is smaller than the symbol size.
		The grayed areas correspond to the positions at which the Efimov states cross the Cs$ + $Cs$ + $Li scattering threshold and their widths represent their uncertainty.}
		\label{fig:loss_rate_energies}
	\end{figure*}	
	
The observed Cs-Cs-Li three-body recombination rates $L_3$ versus the scattering length  $a_\mcsli$ for the broad Cs-Li FR at 843~G at temperatures of 450~nK and 120~nK are shown in Fig.~\ref{fig:loss_rate_energies}(a) together with the theoretical recombination rate from the spinless vdW theory. Here the intraspecies scattering length is in the range $-1560a_0 \lesssim a_\mCsCs \lesssim -1000a_0$. The measured rates of different data sets have been scaled by numerical constants, which are extracted from the universal zero-range theory, but lie well within the experimental uncertainties~\cite{Ulmanis2016}. Three Cs-Cs-Li recombination resonances are evident, while the two excited ones are located in the universal regime [inset of Fig.~\ref{fig:loss_rate_energies}(a)]. The spinless vdW theory~\cite{Wang2012d} is in excellent agreement with the experimental data. The theoretically calculated loss rate spectrum  recovers not only the location of the two excited-state resonances, but also the position of the ground-state resonance. This is in stark contrast to the analysis with the universal zero-range theory~\cite{Petrov2015}, where no agreement between theory and experiment was observed for the ground-state resonance \cite{Ulmanis2016}. Since the essential difference from the zero-range theory is the inclusion of the vdW length scales that determine the short-range behavior of realistic pairwise potentials, we conclude that the previously observed deviation of the CsCsLi ground-state resonance from the universal zero-range theory~\cite{Ulmanis2016} predominantly originates from the vdW interaction. An excellent agreement is also found with the calculated energy spectrum [Fig.~\ref{fig:loss_rate_energies}(b)], where the positions at which the three-body states become unbound [grayed areas in Figs.~\ref{fig:loss_rate_energies}(a) and \ref{fig:loss_rate_energies}(b)] perfectly align with the maxima of the measured recombination rates.

An important question is the extent of the influence of van der Waals forces on the scaling between consecutive Efimov resonances. Therefore, we extract Efimov resonance positions $a_{-}^{(n)}$ and scaling factors $\lambda^{(n)}=a^{(n)}_-/a^{(n-1)}_-$ by three different methods: first, the experimental resonance positions $\mathrm{B}_{\mathrm{expt}}^{(n)}$ are obtained by fitting a Gaussian profile with linear background to the three-body loss rate $L_3(\mathrm{B})$ and conversion to scattering length $a^{(n)}_{-,\mathrm{expt}}$ via the parametrization given in \cite{Ulmanis2015}, where the uncertainty consists of statistical, systematic, and conversion errors. Second, the whole $L_3(\mathrm{B})$ spectrum is fitted with the universal zero-range theory \cite{Petrov2015} and the three-body parameter as well as the inelasticity parameter are extracted. The resonance positions $a^{(n)}_{-,\mathrm{zr}}$ are obtained by setting the temperature and inelasticity parameter in the theory equal to zero~\cite{Ulmanis2016}. By this, finite temperature effects can be eliminated. Third, the calculated trimer energy spectrum from the spinless vdW theory is employed to extract $a^{(n)}_{-,\mathrm{vdW}}$ as the average value of the two numerical grid points, between which the three-body state merges with the scattering continuum [see grayed areas in Fig.~\ref{fig:loss_rate_energies}(b)]. The uncertainty is given by one-half of the step size of the local grid. In this way the influence of finite temperature can be safely neglected, since it modifies the three-body recombination rates, but not the energy spectrum below the scattering continuum. For comparison, the resonance positions $a^{(n)}_-$ and scaling factors $\lambda^{(n)}$ obtained from all three methods are listed in Table~\ref{tab:res_pos}.

\begin{table*}[t]
	\caption{\label{tab:res_pos}
		Cs-Cs-Li Efimov resonance positions $a^{(n)}_-$ and scaling factors $\lambda^{(n)}=a^{(n)}_-/a^{(n-1)}_-$ for the 843~G Cs-Li FR (upper lines) and the 889~G FR (lower lines) obtained by three different methods. The experimental positions $\mathrm{B}_{\mathrm{expt}}^{(n)}$ are extracted by fitting a Gaussian profile with linear background to the three-body loss rate $L_3(\mathrm{B})$ and conversion to scattering length $a^{(n)}_{-,\mathrm{expt}}$ via the parametrization given in \cite{Ulmanis2015}. The values in brackets represent the statistical, systematic, and conversion error. The theoretical values are extracted from the universal zero-range theory \cite{Petrov2015} (843~G FR), the zero-range theory described in Sec.~\ref{sec:zr} (889~G FR) and the spinless vdW model. The values for the vdW model are extracted from trimer energy spectra [see Figs.~\ref{fig:loss_rate_energies}(b) and \ref{fig:loss_rate_energies}(d)] and the quantities in the parentheses represent their uncertainty. Details are given in the text.}
	\begin{ruledtabular}
		\begin{tabular}{cllllllll}
			& $ n $ & $\mathrm{B}_{\mathrm{expt}}^{(n)}$ (G) & $ a^{(n)}_{-,\mathrm{expt}} $  $ (a_0) $ & $ \lambda_{\mathrm{expt}}^{(n)}$ &  $ a^{(n)}_{-,\mathrm{zr}} $ $ (a_0) $ & $ \lambda_{\mathrm{zr}}^{(n)}$ & $ a^{(n)}_{-,\mathrm{vdW}} $ $ (a_0) $ & $ \lambda_{\mathrm{vdW}}^{(n)} $ \\ \hline
			\parbox[t]{2mm}{\multirow{3}{*}{\rotatebox[origin=c]{90}{843 G}}} &
			0   & 848.90(6)(3)\footnote{From \cite{Pires2014} at a temperature of 450~nK.} & -311(3)(1)(1)\footnote{From \cite{Ulmanis2015}.} & - &  -350\footnote{From \cite{Ulmanis2016}.}                    &  -                               &   -317(3)                  & -                         \\
			& 1  & 843.772(10)(16)\footnotemark[3] & -1840(20)(30)(40) & 5.9(1)(1)(1)  &   -1777\footnotemark[3]       & 5.08\footnotemark[3]                            &             -1680(20)  & 5.3(1)    \\
			& 2  & 843.040(10)(16)\footnotemark[3]  & -8140(380)(620)(880) & 4.4(2)(3)(4)  & -9210\footnotemark[3]                   & 5.18\footnotemark[3]                            & -8570(250)                    & 5.1(2)           \\\hline
			\parbox[t]{2mm}{\multirow{2}{*}{\rotatebox[origin=c]{90}{889 G}}} &
			1  & 889.389(4)(16)\footnote{From supplemental material of \cite{Ulmanis2016b}.} & -2130(10)(40)(70)\footnotemark[4]& -  &   -1670       & -                            &             -2150(50)\footnote{From \cite{Ulmanis2016b}.}  & -   \\
			& 2  & 888.787(4)(16)\footnotemark[4] & -8170(160)(620)(890)\footnotemark[4]& 3.8(1)(3)(3)\footnotemark[4]  & -8200                   & 4.91                            & -8500(500)\footnotemark[5]                    & 4.0(3)\footnotemark[5]           \\
		\end{tabular} 
	\end{ruledtabular}
\end{table*}

Remarkably, the experimental and vdW value of $a_-^{(0)}$ agree very well, while the zero-range theory deviates and predicts a much larger value. This deviation can be attributed to short-range effects and was observed as a deviation between the zero-range theory and the measured recombination rates for the ground-state resonance \cite{Ulmanis2016}. Therefore, the scaling factor from the spinless vdW model is larger by about 4\%, if compared to the zero-range theory with actual scattering lengths \cite{Petrov2015}, and larger by about 7\%, if compared to the zero-range theory for noninteracting bosons \cite{Braaten2006,Helfrich2010}. The experiment gives an even larger value for $\lambda^{(1)}$ due to the larger value of $a_{-,\mathrm{expt}}^{(1)}$ in comparison to the zero-range and vdW models.

On the other hand, the vdW and zero-range theory predict a similar scaling factor between the first and second excited Efimov states, highlighting a behavior of the CsCsLi system that is independent of short-range effects.

 \subsection{Positive intraspecies scattering length}
\label{sec:L3pos}

The measured three-body recombination rate spectra close to the 889~G Cs-Li FR ($+180a_0 \lesssim a_\mCsCs\lesssim +360a_0$) at temperatures of 320~nK and 120~nK together with calculated recombination rates from the spinless vdW model are shown in Fig.~\ref{fig:loss_rate_energies}(c). Here, only two Cs-Cs-Li Efimov resonances are observed, while the first recombination resonance is located at $a_\mcsli\approx -2000a_0$. This is about a factor of seven larger than in the case of negative $a_\mCsCs$ [see Fig.~\ref{fig:loss_rate_energies}(a)], which is consistent with the energy spectrum of the Efimov states [see Fig.~\ref{fig:loss_rate_energies}(d)], where the most deeply bound state predissociates into a universal atom-dimer state before reaching the three-body continuum~\cite{Ulmanis2016b} and hence does not generate a resonance at the scattering threshold. This is consistent with the findings of Sec.~\ref{sec:zr}, where the existence of a Feshbach dimer $BB$ at positive intraspecies scattering lengths $a_{BB}>0$ leads to a splitting into two Efimov branches asymptotically connecting to the three-atom and atom-dimer channels. Therefore, we assign the first recombination feature to the first excited Efimov resonance.  

The calculated recombination spectra from the vdW model show qualitative agreement with the experimental observations. However, a shift between the experimental and theoretical resonance positions will require more detailed analysis and may be due to the multichannel nature of the employed FR \cite{Wang2014a,Sorensen2012} for tuning $a_\mcsli$. It is shown that significant deviation of the Efimov resonance positions from the spinless vdW theory has been observed near a FR with $s_{\mathrm{res}}$ about 35 times smaller than those in our current cases~\cite{Johansen2017}. However, the  resonance positions obtained from the energy spectra (grayed areas) coincide very well with the experimental observations. Within the framework of an effective field theory corrections due to finite effective range and intraspecies scattering length on three-body recombination rates and Efimov resonance positions have been studied recently \cite{Acharya2016}.

Similar to Sec.~\ref{sec:L3neg} we extract resonance positions and scaling factors by three different methods. The experimental and vdW parameters are extracted as described previously. Here,  $a^{(n)}_{-,\mathrm{zr}}$ are calculated within the zero range model described in Sec.~\ref{sec:zr}. We find the energy levels of the upper branch adiabatic hyperspherical potential curves (see Fig.~\ref{fig:Cs2Li_Pot2}) and search for the values of $a_\mcsli$ where their energy intersects zero. As opposed to the negative intraspecies scattering length case, there is no three-body parameter necessary. The positions and scaling factors are listed in Table~\ref{tab:res_pos}.

The experimentally determined position of the first excited Efimov resonance $a_-^{(1)}$ close to the 889~G FR is shifted by $\approx 300a_0$ with respect to the 843~G FR, while the position of the second excited resonance $a_-^{(2)}$ is nearly unchanged. The resonance position $a_-^{(2)}$ extracted by all three methods agree very well with each other, while for the first excited state the zero-range theory clearly deviates from the two other models. This deviation may be explained by neglect of nonadiabatic couplings in our single-channel zero-range theory. Additionally, the binding energy of the Cs$_2$ dimer is not reproduced in this model by disregard of effective range corrections. However, the quantitative influence of these limitations requires further detailed analysis. The scaling factor obtained from the zero-range theory agrees very well with the theoretically predicted scaling factor of 4.9 between consecutive CsCsLi Efimov resonances \cite{Braaten2006,Helfrich2010a,Helfrich2010}. The obtained scaling factors from experiment and vdW theory of 3.8(1)(3)(3) and 4.0(3) between the two excited Efimov resonances clearly deviate from the universal zero-range theory.

\subsection{Scaling laws of three-body recombination rates}
\label{sec:powerlaws}

\begin{figure}[t]
	\centering
	\includegraphics[width=1\linewidth]{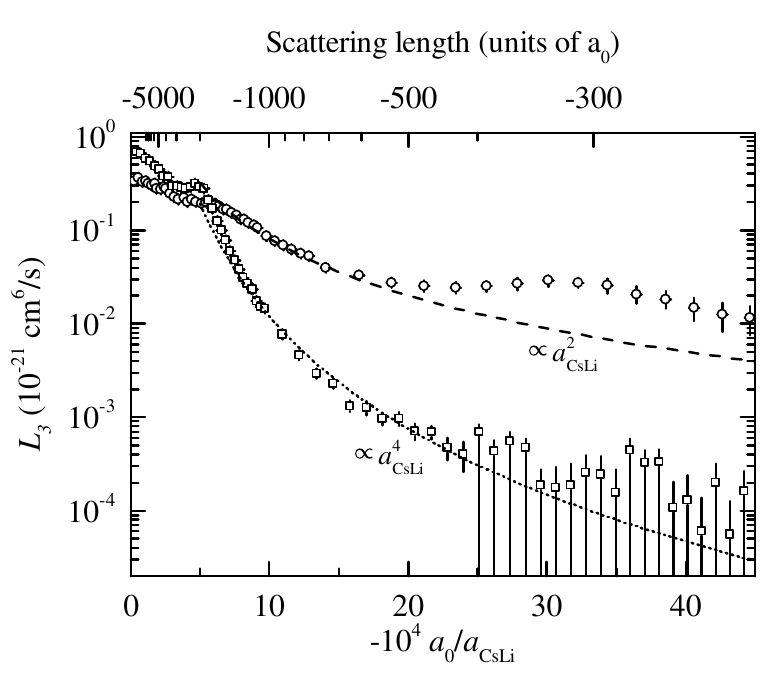}
	\caption{Cs-Cs-Li three-body recombination rate spectra with positive (squares) and negative (circles) intraspecies interaction. The power laws $\propto a_{\mcsli}^4$ (dotted line) and $\propto a_{\mcsli}^2$ (dashed line) are drawn as guides to the eye. The error bars represent the statistical errors from bootstrapping, magnetic field uncertainties, and technical limit due to Cs-Cs-Cs three-body losses. The data were taken close to the 889~G Cs-Li Feshbach resonance at a temperature of 320~nK, where $ a_\mCsCs\approx+190a_0 $, and close to the 843~G Cs-Li Feshbach resonance at a temperature of 450~nK, where $ a_\mCsCs\approx-1500a_0 $. }
	\label{fig:power_laws}
\end{figure}

The scaling behavior of three-body recombination in a heteronuclear system is expected to drastically depend on the sign and magnitude of the intra- and interspecies scattering lengths~\cite{DIncao2009a}. We study this behavior by comparing the three-body recombination spectra close to the 843~G and 889~G Cs-Li FRs for temperatures of 450~nK and 320~nK in Fig.~\ref{fig:power_laws}. They feature power-law scaling behavior with the scattering lengths $ a_\mcsli $ and $ a_\mCsCs $, which qualitatively agree with the expected scaling laws near overlapping Feshbach resonances~\cite{DIncao2009a}. In certain ranges of $a_\mcsli$ each of the loss rate spectra corresponds to one of two distinctive cases of three-body scattering: the one near the 889~G FR can be characterized by $ \left| a_{\mcsli}\right| \gg a_\mCsCs\approx 190a_0$ for which $ L_3 \propto a_{\mcsli}^4 $ is expected, whereas the one near the 843~G FR for small $ a_{\mcsli} $ approximately corresponds to the case of $ \left| a_{\mcsli}\right| \ll \left|  a_{\mathrm{Cs}} \right| \approx 1500a_0 $ with an expected scaling of $ L_3 \propto a_{\mcsli}^2 a_\mCsCs^2 $~\cite{DIncao2009a}. The scaling laws can be explained by tunneling through effective three-body potential barriers within a simple WKB model. The power laws are displayed in the respective range in Fig.~\ref{fig:power_laws} as a guide to the eye. Since the experimentally employed scattering lengths only approximately capture the inequalities imposed by the theory, especially in the latter case, the power laws can only approximately recover the behavior of the actual Cs-Li system. Close to the pole of the FR the theoretical scaling does not apply anymore due to the unitarity limit.

The shown power law for the 843~G FR does not account for varying $ a_\mCsCs $, which is tuned simultaneously with the magnetic field and changes by approximately a factor of 1.5 for the experimentally employed Feshbach resonance (see Fig.~\ref{fig:scatteringlengths}). 	
For example, in the case of the 843~G Cs-Li Feshbach resonance, far away from the resonance the inequality $ \left| a_{\mcsli}\right| \ll \left|  a_\mCsCs \right|$ is fulfilled, whereas close to the pole of the resonance, the opposite is true, i.e., $ \left| a_{\mcsli}\right| \gg \left|  a_{\mCsCs} \right|$. If $ a_\mCsCs $ was a constant, this would lead to a qualitative change in the power law from $L_3\propto a_\mcsli^2 a_\mCsCs^2$ to $L_3\propto a_\mcsli^4 $ for small to large scattering lengths, respectively. Such a transition in the present data is masked by finite-temperature and short-range effects. However, it might become observable in samples with further reduced temperature. Similar behavior can be found in the universal zero-range theory with finite intraspecies scattering length~\cite{Petrov2015} and formalisms based on optical potentials~\cite{Mikkelsen2015}.
	
The different observed power laws enable us to manipulate the three-body loss in a heteronuclear system. By choosing an appropriate FR, we can control the intraspecies scattering length between the heavy bosons and by this drastically influence the three-body loss rate. For scattering lengths $a_\mcsli\lesssim 500a_0$ the Cs-Cs-Li losses are reduced by approximately two orders of magnitude when changing from large and negative to small and postive intraspecies scattering lengths, paving the way to produce long-lived strongly interacting Bose-Fermi mixtures.
	
\section{Conclusion}
	
In summary, we have presented three theoretical methods at different levels of complexity to solve the three-body Schr\"odinger equation for two heavy identical bosons and one distinguishable particle and compared them to measurements of three-body recombination rates in a system of ultracold Cs and Li atoms. By detailed analysis we confirm the decisive influence of the intraspecies scattering length on the heteronuclear Efimov effect.

The minimalistic hybrid Born-Oppenheimer model gives an intuitive understanding of the strong dependence of the energies on the intraspecies scattering length. Within this framework the binding energies of the three-body states for $a_\mcsli\rightarrow \infty$ in dependence of the intraspecies scattering length $a_\mCsCs$ were calculated and a steplike behavior close to the Cs-Cs vdW length $|a_\mCsCs|\approx r_{\mathrm{vdW}}^{\mCsCs}$ was predicted, analogous to calculations in the hyperspherical vdW model~\cite{Wang2012d}, qualitatively explaining the experimentally observed change in the resonance position of the first excited Efimov state $a_-^{(1)}$ between the two Cs-Li FRs.

Within the hyperspherical adiabatic zero-range theory two qualitatively distinct cases of the heteronuclear Efimov scenario were shown. While for $a_{BB}<0$ the original Efimov scenario is observed, with an infinite number of log-periodically spaced three-body states, the existence of a weakly bound dimer for $a_{BB}>0$ splits the adiabatic hyperspherical potentials into two Efimov branches. Efimov states situated in the lower potential branch may not lead to Efimov resonances in the $B+B+X$ channel as observed in our measured three-body recombination rate spectra. For the upper branch a universal potential barrier at $R\approx2a_{BB}$ makes the introduction of an artificial three-body parameter superfluous and allows for the determination of Efimov resonance positions. However, the applied adiabatic approximation and the neglect of effective range corrections may affect the observed deviations from the experimentally determined resonance positions.

An encouraging level of understanding of the heteronuclear Efimov effect is provided by the spinless van der Waals theory in the adiabatic hyperspherical approximation, for both positive and negative intraspecies scattering lengths. This theory is compared to our measurements of three-body recombination rate spectra in an ultracold mixture of Cs and Li atoms at temperatures as low as 120~nK for two Cs-Li FRs, characterized by different sign and magnitude of the intraspecies scattering length $a_\mCsCs$. We find excellent agreement between experiment and theory on the negative side of $a_\mCsCs$. For positive intraspecies interactions a good agreement between the observed resonances and the energy spectrum is observed. However, a shift between experimental and theoretical three-body recombination rates demands further analysis and may be due to the multichannel character of the Feshbach resonance. No Efimov resonance which can be assigned to the ground Efimov state is observed at the Cs+Cs+Li threshold for positive $a_\mCsCs$. Before reaching the three-body continuum the most deeply bound state dissociates at the atom-dimer threshold.

Away from the two Cs-Li FRs we observe power-law scalings of the three-body recombination rates, which can be attributed to two distinct cases of overlapping Feshbach resonances. We are able to suppress the three-body loss by up to two orders of magnitude via choosing a different Cs-Li Feshbach resonance and drastically increase the lifetime of the Bose-Fermi mixture. A predicted qualitative change in the scaling behavior of the three-body recombination rate close to the 843~G FR from $L_3\propto a_\mcsli^2 a_\mCsCs^2$ to $L_3\propto a_\mcsli^4$ might be observable in mixtures with further reduced temperature.

\vspace*{0.5cm}
	\begin{acknowledgments}
		We are grateful to P. Giannakeas, C. Zimmermann, D. Petrov, R. Grimm, and F. Ferlaino for fruitful discussions. This work is supported in part by the Heidelberg Center for Quantum Dynamics. S.H. acknowledges support by the IMPRS-QD. E.D.K. is indebted to the Baden-W\"urttemberg Stiftung for the financial support of this research project by the Eliteprogramme for Postdocs. Y.W. acknowledges support by an AFOSR-MURI and Department of Physics, Kansas State University, and C.H.G. from the Binational Science Foundation, Grant No. 2012504 and the U.S. National Science Foundation Grant No. PHY-1607180. This work was supported in part by the Deutsche Forschungsgemeinschaft under Project No. WE2661/11-1 and Collaborative Research Centre "SFB 1225 (ISOQUANT)".
	\end{acknowledgments}

\end{document}